\documentclass[twocolumn,showpacs,preprintnumbers]{revtex4}

\input{epsf}

\def\AJ{{\it Ap. J.} }

\def\AJS{{\it Ap. J. Supp.} }

\def\ASAS{{\it Astron. and Astrophys.} }

\def\CQG{{\it Class. Quantum Gravity} }

\def\IJMP{{\it Int. J. Mod. Phys.} }

\def\MPL{{\it Mod. Phys. Lett.} }
\def\MNRAS{{\it Mon. Not. R. Ast. Soc.} }
\def\NAT{{\it Nature} }

\def\PL{{\it Phys. Lett.} }
\def\PR{{\it Phys. Rev.} }
\def\PRL{{\it Phys. Rev. Lett.} }

\def\al{\alpha}  \def\ga{\gamma} \def\de{\delta}
\def\ep{\epsilon}   
\def\th{\theta}   \def\ka{\kappa}
\def\la{\lambda}   
\def\si{\sigma}   
   
  \def\De{\Delta} \def\Th{\Theta}
\def\La{\Lambda}   
 \def\Om{\Omega} \def\mn{{\mu\nu}}

 \def\frac#1#2{{\textstyle{{#1}\over
{#2}}}} 
\def\lsim{\mathrel{\rlap{\lower4pt\hbox{\hskip1pt$\sim$}}
\raise1pt\hbox{$<$}}}
\def\gsim{\mathrel{\rlap{\lower4pt\hbox{\hskip1pt$\sim$}}
\raise1pt\hbox{$>$}}} \def\sqr#1#2{{\vcenter{\vbox{\hrule height.#2pt
\hbox{\vrule width.#2pt height#1pt \kern#1pt \vrule width.#2pt} \hrule
height.#2pt}}}}
\def\square{\mathchoice\sqr66\sqr66\sqr{2.1}3\sqr{1.5}3}
\def\beq{\begin{equation}} \def\eeq{\end{equation}}
\def\beqa{\begin{eqnarray}} \def\eeqa{\end{eqnarray}}

\begin{document}

\title{Do $f(R)$ theories matter?}

\vskip 0.2cm

\author{O. Bertolami}
\email{orfeu@cosmos.ist.utl.pt}

\author{J. P\'aramos}
\email{jorge.paramos@ist.utl.pt}

\vskip 0.2cm

\affiliation{Instituto Superior T\'ecnico, Departamento de
F\'{\i}sica\footnote{Also at  {\it Instituto de Plasmas e Fus\~ao Nuclear, IST}, Lisbon}, \\Av. Rovisco Pais 1, 1049-001 Lisboa, Portugal}

\vskip 0.2cm

\vskip 0.5cm

\date{\today}

\begin{abstract}
We consider a modified action functional with a non-minimum coupling between the scalar curvature and the matter Lagrangian, and study its consequences on stellar equilibrium. Particular attention is paid to the validity of the Newtonian regime, and on the boundary and exterior matching conditions, as well as on the redefinition of the metric components. Comparison with solar observables is achieved through numerical analysis, and constraints on the non-minimum coupling are discussed.

\vskip 0.5cm

\end{abstract}

\pacs{04.20.Fy, 04.80.Cc, 97.10.Cv \hspace{2cm}Preprint DF/IST-7.2007}

\maketitle


\section{Introduction}

Modern cosmology faces two outstanding challenges, namely the existence and nature of dark energy and dark matter. Many theories have been put forward to address both issues: for dark matter, several candidates are available, such as weak-interacting particles (WIMPs) arising from extensions to the Standard Model ({\it e.g.} axions, neutralinos), {\it etc.}; for dark energy, ``quintessence'' models consider the slow-roll of a scalar field \cite{scalar, Amendola}, amongst other candidates; others suggest that the averaging of inhomogeneities at a cosmological scale may yield an effective scalar field, thus accounting for the dark energy component of the Universe \cite{inhomo}. A possible unification of both ``dark'' components has also been suggested, resorting to a scalar field model \cite{Rosenfeld} or an exotic equation of state, as featured by the so-called modified Chaplygin gas \cite{Chaplygin}.

A different approach assumes that no extra energy content is needed and that the fundamental laws and tenets of gravitation may be incomplete, perhaps just a low-energy approximation; as a consequence, modifications of the Friedmann equation to include higher order terms in the energy density $\rho$ (see {\it e.g.} \cite{Maartens} and references therein) have been proposed or, at a more fundamental level, changes to the action functional. A rather straight forward approach lies in replacing the linear scalar curvature term in the Einstein-Hilbert action by a function of the scalar curvature, $f(R)$; alternatively, one could resort to other scalar invariants of the theory \cite{f(R)}. This has led to some success in replicating the accelerated expansion of the Universe, while comparison with its evolution throughout the different ages has yielded some constraints and exclusions to the form of $f(R)$; perhaps the most well-known proposal of this type is the Starobinsky inflationary model, where a quadratic term in the curvature is added to the usual linear form (plus cosmological constant), $f(R)=R - \La + \al R^2$ \cite{Staro}. Solar system tests could also bring further insight, mostly arising from the parameterized post-Newtonian (PPN) metric coefficients derived from this extension of general relativity (GR). However, some disagreement exists in the community, with some arguing that no changes are predicted at a post-Newtonian level (see {\textit e.g.} \cite{PPN} and references therein); amongst other considerations, this mostly stems from an approach based either in the more usual metric affine connection (that is, where the affine connection is taken {\it a priori} as depending on the metric), or in the so-called Palatini approach \cite{Palatini} (where both the metric and the affine connection are taken as independent variables). As an example of a clear phenomenological consequence of this extension of GR, it has been shown that $f(R) = f_0 R^n$ theories yield a gravitational potential which displays an increasing, repulsive contribution, added to the Newtonian term \cite{flat}.

Notwithstanding the significant literature on these $f(R)$ models, few steps have been taken to address another interesting possibility: not only that the curvature is non-trivial in the Einstein-Hilbert Lagrangian, but also that the coupling between matter and geometry is not minimum; indeed, these are only implicitly related in the action functional, since one expects that covariantly invariant terms in $\mathcal{L}_m$ should be constructed by contraction with the metric ({\it e.g.} the kinetic term of a real scalar field, $g^\mn \phi_{,\mu} \phi_{,\nu}$). A non-minimum coupling would imply that geometric quantities (such as the scalar invariants) would explicitly show in the action; asides from theoretical elegance, this could have deep phenomenological implications: indeed, in regions where the curvature is high (which, in GR, are related to regions of high energy density or pressure), the implications of such theory could deviate considerably from those predicted by Einstein's theory \cite{Lobo}. Related proposals have been put forward previously to address the problem of the accelerated expansion of the Universe \cite{expansion} and the existence of a cosmological constant \cite{cosmological}.

In this sense, the immediate question, posed in the title of this work, is: what are the implications for the behaviour of matter under such conditions? Some work has been put forward concerning this issue, namely changes to geodetic behaviour \cite{Lobo}, the possibility of modelling dark matter \cite{dark matter} and the violation of the highly constrained equivalence principle \cite{equivalence}. Perhaps the most important consequence of the mentioned studies is that energy may no longer be covariantly conserved, that is, $\nabla^\mu T_\mn \neq 0$, where $T_\mn$ is the energy-momentum tensor of matter; this occurs because, due to the presence of extra terms in the equations of motion, the Bianchi identities no longer imply in the covariant conservation of the energy-momentum tensor.

This work addresses what we believe is the natural proving ground for a non-minimally coupled gravity model: regions where the density may be high enough, to evidence some deviation from GR, although moderate enough so that effects are still perturbative -- a star. The results rely upon and expand the methodology followed by the authors in previous studies \cite{astro,darkstar}. This paper is divided in the following sections: first, we present the model upon which the subsequent work is based; then, we develop the equations of motion, aiming at the modified Tolman-Oppenheimer-Volkoff (TOV) equation, with due care taken regarding the validity of the Newtonian regime and resulting modified hydrostatic equilibrium equation; afterwards, we insert the polytropic equation of state into the latter, and compute the necessary observables; a numerical session follows, where profiles and bounds are computed for the relevant quantities; finally, a discussion of our results is presented.

\section{The model}

Following the discussion of the previous section, one postulates the following action for the theory \cite{Lobo},

\beq S = \int \left[ {1 \over 2}f_1(R) + [1 + \la f_2(R) ] \mathcal{L}_m \right] \sqrt{-g} d^4 x \label{action} \eeq

\noindent where $f_i(R)$ (with $i=1,2$) are arbitrary functions of the scalar curvature $R$, $\mathcal{L}_m$ is the Lagrangian density of matter and $g$ is the metric determinant. For convenience, the contribution of the non-minimum coupling of $f_2$ is gauged through the coupling constant $\la$ (which has dimensions $[\la] = [f_2]^{-1}$). The standard Einstein-Hilbert action is recovered by taking $f_2=0$ and $f_1= 2 \ka (R - 2 \La)$, where $\ka = c^4 /16 \pi G$ and $\La$ is the cosmological constant (from now on, one works in a unit system where $c= 1$).

Variation with respect to the metric $g_\mn$ yields the modified Einstein equations of motion, here arranged as

\beqa && \label{EE} \left( F_1 + 2 \la F_2 \mathcal{L}_m \right) R_\mn - {1 \over 2} f_1 g_\mn = \\* \nonumber && \left( \square_\mn - g_\mn \square \right) \left(F_1 + 2 \la F_2 \mathcal{L}_m \right) + \left[ 1 + \la f_2 \right] T_\mn~~, \eeqa

\noindent where one defines $\square_\mn \equiv \nabla_\mu \nabla_\nu$ for convenience, as well as $F_i(R) \equiv f'(R)$, and omitted the argument. The matter energy-momentum tensor is, as usually, defined by

\beq T_\mn = -{2 \over \sqrt{-g}} {\de \left(\sqrt{-g} \mathcal{L}_m \right) \over \de g^\mn } ~~. \eeq

As stated before, the Bianchi identities, $\nabla^\mu G_\mn = 0$ imply the non-(covariant) conservation law

\beq \nabla^\mu T_\mn = {\la F_2 \over 1+ \la f_2} \left( g_\mn \mathcal{L}_m - T_\mn \right) \nabla^\mu R ~~, \label{non-cons} \eeq

\noindent and, as expected, in the GR limit $\la \rightarrow 0$, one recovers the conservation law $\nabla^\mu T_\mn = 0$.

\subsection{Scope of application}

It is our stated purpose to arrive at the modified form of the TOV equation, the relativistic version of the hydrostatic equilibrium condition. From Eq. (\ref{EE}), it is clear that a full treatment of the equations of motion is unattainable unless some specific form for $f_1(R)$ and $f_2(R)$ is provided. Furthermore, the presence of both the pure curvature and non-minimum coupling, respectively, still constitutes a daunting analytical challenge; hence, and since one is mainly interested in the relevance of the effects within a high curvature and pressure medium, where $f_2$ should overwhelm the modification of the pure curvature term, $f_1 - 2 \ka R $ (neglecting the contribution of the cosmological constant), it appears sensible to discard the latter; a treatment of the standard $f(R)$ scenario with $f_2=0$ may be found in Ref. \cite{TOV}. 

Mathematically, the chosen approximation reads as

\beqa \label{ineq1} {1 \over 2}| f_1 - \ka R| & \ll & |\la f_2 \mathcal{L}_m |~~, \\* \nonumber |F_1 - 2 \ka| & \ll & 2 | \la F_2 \mathcal{L}_m| ~~, \\* \nonumber \left| \left( \square_\mn -g_\mn \square \right) F_1 \right| & \ll & 2 \left| \la \left( \square_\mn -g_\mn \square \right) (F_2 \mathcal{L}_m) \right| ~~. \eeqa

\noindent where the second and third inequalities indicate that this regime stems not just from the comparison between contributions to the action functional, but also those involved in the modified Einstein equations (\ref{EE}); also, notice that the perturbative condition $\la f_2(R) \ll 1$ has not been enforced yet. The last inequality is satisfied if the following stronger conditions hold,

\beqa && \left| {d F_1 \over dr} \right| \ll 2\left| \la{d(F_2\mathcal{L}_m) \over dr} \right| ~~, \\* \nonumber && \left| {d^2 F_1 \over dr^2} \right|\ll 2 \left| \la {d^2(F_2 \mathcal{L}_m) \over dr^2} \right|~~. \eeqa

This said, one considers the simplest form $f_2 =R $; this linear coupling could arise from a first order expansion of a more general $f_2(R)$ function, in the weak field environ of the Sun. Also, one takes $\mathcal{L}_m = p$, a natural choice for the Lagrangian density of an ideal fluid \cite{Lagrangian} -- characterised by the standard energy-momentum tensor

\beq T_\mn = (\rho + p) u_\mu u_\nu - p g_\mn ~~, \label{Tmn} \eeq

\noindent where $p $ is the fluid's pressure, $\rho$ its matter-energy density and $u$ its 4-velocity vector, with $u_\mu = (u_0,\vec{0})$, $u_\mu^\mu = 1$, so that $u_0 = g_{00}^{1/2}$. Inserting the above expressions for $f_2$ and $\mathcal{L}_m$, the inequalities (\ref{ineq1}) become

\beqa \label{ineqa} |f_1 | & \ll & 2 |\left(\ka + \la p \right) R |~~, \\* \label{ineqb} | F_1 | & \ll & 2 \left| \ka + \la p \right|~~, \\* \label{ineqc} \left| {d F_1 \over dr} \right| & \ll & 2 \left| \la p'(r) \right|~~~,~~~\left| {d^2 F_1 \over dr^2}\right| \ll 2 | \la p''(r) | ~~,\eeqa

\noindent where the prime denotes derivation with respect to the radial coordinate. Clearly, if $f_1 = 2 \ka R \rightarrow F_1 = 2\ka$, these are trivially satisfied. Also, notice that $f_2 = R$ implies that $[\la] = [f_2^{-1}]=[R]^{-1} =M^{-2}$.

One must now ascertain the form of $f_1(R)$. In Ref. \cite{Amendola} it is shown that acceptable models with ${f_{1a} = 2 \ka( R - \al R^{1-m})}$ ($\al > 0 $, ${[\al] = [R^m] = M^{2m}}$ and $0 < m < 1 $ or ${f_{1b} = 2\ka( R + \al R^2 - \La)}$ (with $\al \La \ll 1 $ and ${[\al] = [R^{-1}]= M^{-2}}$) are cosmologically viable; the latter may arise from the renomalizability of the theory near the Planck scale, with $\al \sim M^{-2}$ and $M = 10^{12}~GeV$ \cite{Amendola}. For the perturbative regime to be valid, one must have, for the $f_1 = f_{1a}$ case

\beqa && \label{ineqf1a} |\la| \gg \al \ka\left|{1 \over R^m p}\right| ~~, \\* \nonumber && |\la| \gg (1-m) m \al \ka \left| {1 \over R^{1+m}p'}{d R \over dr} \right| ~~, \\* \nonumber && |\la | \gg (1-m)m \al \ka \left|{1 \over R^{1+m}p''} \right| \left| {d^2 R\over dr^2} -(1+m){\left({d R \over dr}\right)^2\over R} \right| ~~.\eeqa

\noindent while, for the $f_1=f_{1b}$ case (which can be derived from the above inequalities, setting $m=-1$, $\al \rightarrow -\al$ and ignoring the cosmological constant term), 

\beqa && \label{ineqf1b} |\la| \gg \al \ka\left|{R \over p}\right| ~~, \\* \nonumber && |\la| \gg 2 \al \ka \left| {1 \over p'}{d R \over dr} \right| ~~, \\* \nonumber && |\la | \gg 2 \al \ka \left|{1 \over p''} \right| \left| {d^2 R\over dr^2} \right| ~~.\eeqa

\subsection{Equation of motion}

With the above choice for $f_1$ and $f_2$, the equation of motion becomes

\beqa && \left( 1 + a p \right) R_\mn - {1 \over 2}R \left( g_\mn +a T_\mn \right) = \\* \nonumber && a (\square_\mn - g_\mn \square)p + {1 \over 2 \ka}T_\mn ~~, \eeqa

\noindent where one defines the parameter $a \equiv \la / \ka = 16\pi G \la $, with dimension $[a ] = M^{-4}$; accordingly, both $ap$ and $a\rho$ are dimensionless quantities. In the above form, the physical meaning of the proposed model is more transparent: aside from a pressure-dependent term on the {\it r.h.s.}, the most interesting modification occurs on the {\it l.h.s.}; firstly, the contribution of the Riemann tensor is modified by a factor $1 + a p$; secondly, the scalar curvature is coupled not only to the metric $g_\mn$, but also to the energy-momentum tensor $T_\mn$. One thus gets a clear picture of the matter-geometry interaction, which occurs via scalar-tensor combinations.

By taking the trace of the above equation, one obtains

\beq R = - {-3a \ka \square p + T \over \ka \left[2+a \left(T-2p \right)\right]} = {3p - \rho + 3 a \ka \square p \over \ka \left[2+a(\rho - 5 p ) \right]} ~~, \label{ricci} \eeq

\noindent having inserted $T = T_\mu^\mu = \rho - 3p$. Interestingly enough, in the ``strong'' $a \rightarrow \infty$ regime it yields the asymptotic result $R = 3 \square p / (\rho - 5p) $: in this regime, a varying, low density can give origin to extremely high curvatures, while an almost uniform, even if a high density fluid might yield a vanishingly small curvature. 

\subsection{Perturbative regime}

Foreseeing a later comparison with solar observables, which one knows are well predicted by GR, it is now assumed that the effect of $f_2$ yields only perturbative corrections: in the approximation $|\la f_2| = |\la R| \ll 1$, the scalar curvature is given by $R \approx (3p-\rho)/\ka$, resulting in $|(\la / \ka) (3p - \rho) |= |a (3p - \rho)| \ll 1$; anticipating the Newtonian approximation $p \ll \rho$, this yields $|a| \rho \ll 1 $ and $|a| p \ll 1$. Inserting Eq. (\ref{ricci}) in the Einstein equation, one obtains, after some algebraic manipulation,

\beqa \label{motion} && \ka [2 + a (\rho - 3 p )] R_\mn = \\* \nonumber && (3p - \rho) g_\mn +2 (1-ap) T_\mn + a \ka (4 \square_\mn - g_\mn \square) p ~~, \eeqa

\noindent keeping only first order terms in $a$. This shall be the main tool for the derivation of the TOV equation. 

\section{Stellar equilibrium}

\subsection{Static, spherical symmetric scenario}

Since one is dealing with an ideal, spherically symmetric system, in which temporal variations are assumed to occur only at the cosmological scale $H_0^{-1}$, and hence negligible at an astrophysical time scale, one considers the Birkhoff metric (in its anisotropic form), given by the line element

\beq ds^2 = e^{\nu(r)} dt^2 - \left( e^{\si(r)}dr^2 +d \Om^2 \right)~~, \label{line} \eeq

\noindent with $d \Om = r^2 (d \th^2 + sin^2 \th ~d \phi^2)$, so that $\sqrt{-g} = r^2 ~sin\th~e^{(\nu + \si)/2}$.

One can resort to the expression of the Riemann tensor and Eq. (\ref{motion}) to obtain the following intermediary step (the full algebraic derivation is shown in Appendix A): 

\beqa \label{eq1} && \ka [2 + a (\rho - 3 p )] \left[ {1 \over r^2 }+ e^{-\si} \left({\si' \over r} - { 1\over r^2 }\right) \right] = \rho - \\* \nonumber && {ap\over 2}(\rho+3p) + {a \ka \over 2} \left( 5 e^{-\nu} \square_{00} + 3e^{-\si} \square_{rr} + 2{ \square_{\th\th} \over r^2} \right) p~~. \eeqa 

One now defines the parameter $m_e$, here called the effective mass, to distinguish it from its identification with the gravitational mass of the unperturbed GR scenario, derived from the Schwarzschild metric; it is given by the usual expression

\beq e^{-\si} = 1 - {2 Gm_e \over r}~~, \label{massdef} \eeq

\noindent which yields

\beq \left[ {1 \over r^2} + e^{-\si} \left({\si' \over r} - { 1\over r^2 }\right) \right] = {2Gm_e' \over r^2}~~, \eeq

\noindent where the prime denotes differentiation with respect to the radial coordinate. Inserting the above in Eq. (\ref{eq1}) and solving for $m'_e$, one gets

\beqa && m'_e = 4 \pi r^2 \rho {2 - ap \left(1 + {3p \over \rho} \right) \over 2 + a (\rho - 3 p )} + \\* \nonumber && {a r^2 \over 4 G} {\left( 5 e^{-\nu} \square_{00} + 3e^{-\si} \square_{rr} + {2 \over r^2} \square_{\th\th} \right) p \over 2 + a (\rho - 3p )} ~~. \eeqa

\noindent In the perturbative regime, one may take only first-order terms in $ap$ and $a\rho$, obtaining

\beqa \label{mass'} && m'_e \approx 4 \pi r^2 \rho \left[ 1 + a \left(p - {\rho \over 2} - {3 \over 2} {p^2 \over \rho} \right) \right] + \\* \nonumber && {a r^2 \over 8 G} \left( 5 e^{-\nu} \square_{00} + 3e^{-\si} \square_{rr} + 2{ \square_{\th\th} \over r^2} \right) p ~~, \eeqa 

\noindent using $4G\ka = 1/4\pi$. The above expression clearly shows the perturbation to the purely gravitational mass, defined by $m'_g = 4 \pi r^2 \rho$.

\subsection{Newtonian limit}
 
 Before continuing the derivation, it is opportune to address the issue of the validity of the Newtonian regime; clearly, establishing this limit will enable many fruitful simplifications in the calculations ahead. In the standard derivation of the hydrostatic equilibrium equation, this arises from a set of simplifications imposed on the relativistic TOV equation; the later reads,
 
 \beq p'(r) = -{G \over r^2} {[\rho(r) + p(r) ] [ m_e(r) + 4 \pi p(r) r^3 ] \over 1 - 2 Gm_e(r)/r}~~, \label{TOVGR} \eeq
 
 \noindent and the Newtonian approximation is valid if the following inequalities are satisfied
 
 \beq r \gg 2Gm_e(r)~,~ \rho(r) \gg p(r)~,~ m_e(r) \gg 4 \pi p(r) r^3 ~~, \label{newtonGR} \eeq
 
 \noindent yielding the non-relativistic hydrostatic equation of state
 
 \beq p'(r) = - G {\rho(r) m_e(r) \over r^2 }~~. \label{hydroGR} \eeq

Since is is assumed that the coupling between matter and geometry will only produce a perturbative effect, leading to the redefinition of mass through Eq. (\ref{mass'}), it is clear that no changes occur regarding the validity of the Newtonian approximation. Hence, foreseeing the application of the following results to the Sun, where such regime is valid, one may simplify the intermediate calculations and, when convenient, insert the inequalities (\ref{newtonGR}) in order to simplify $a$-dependent terms. By the same token, terms involving the coupling between $a$-dependent quantities and any covariant derivatives may be evaluated by taking their Newtonian counterparts,

\beqa \label{derNewt} && \square_{00} p = -{e^{\nu - \si} \over 2}\nu' p' \approx 0~~, \\* \nonumber && \square_{rr}p = p'' + {1 \over 2}\si' p' \approx p''~~, \\* \nonumber && \square_{\th\th} p = e^{-\si}rp' \approx rp'~~. \eeqa

\noindent Notice that this does not imply that one is directly deriving the Newtonian hydrostatic equilibrium equation, since leading order terms will not be approximated.

\subsection{Tolman-Oppenheimer-Volkoff equation}

By following a procedure similar to the one leading to Eq. (\ref{eq1}) (depicted in Appendix A), one obtains the following equation:

\beqa \label{eq2} && \ka \left[2 + a \left(\rho - 3 p \right) \right] \left[ \left(1 - {2Gm_e \over r}\right) { \nu' \over r } -{2 G m_e \over r^3} \right] = \\* \nonumber && p +{ap \over 2} (p-\rho ) + {a \ka \over 2} \left( 3 e^{-\nu} \square_{00} + 5 e^{-\si} \square_{rr} - 2 {\square_{\th\th} \over r^2 } \right) p ~~. \eeqa 

\noindent Substituting by the expressions for the covariant derivatives, one gets

\beqa && \left( 3 e^{-\nu} \square_{00} + 5 e^{-\si} \square_{rr} - 2 {\square_{\th\th} \over r^2 } \right) p = \\* \nonumber && 5 \left( 1 - {2 G m_e \over r }\right)p'' + \\* \nonumber && \left[ 5 { G m'_e \over r } - \left( 1 - {2G m_e\over r} \right) {3 \nu' \over 2} - {1 \over r } \left(2 + {Gm_e \over r} \right) \right] p'~~.\eeqa

\noindent One now introduces the approximations discussed in the previous subsection. Also, the approximation $ m'_e \approx m'_g = 4 \pi \rho r^2$ is taken, since the perturbative corrections would produce second order terms in $a$. One obtains, after a little algebra,

\beqa && \ka \left( 2 + a\rho \right) \left( { \nu' \over r } -{2 G m_e \over r^3} \right) = \\* \nonumber && p \left(1 - {a \over 2} \rho \right) + {a \ka \over 2} \left[ 5 p'' + \left( 5 { G m'_e \over r } -{3 \nu' \over 2} - {2 \over r } \right) p' \right] ~~. \eeqa 

\noindent Solving for $\nu'$, one gets

\beqa {\nu' \over 2} =&& G{ \left( 2 + a \rho \right) m_e + 4 \left(2 - a \rho \right) \pi p r^3 \over r \left(r - 2 G m_e\right)\left[2 + a\left( \rho + {3 \over 4} r p'\right) \right]} + \\* \nonumber && a {r^2 \over r - 2Gm_e} { {5 \over 4} p'' + \left( 5 \pi G \rho r - {1 \over 2 r }\right) p' \over 2 + a\left( \rho + {3 \over 4} r p'\right) } ~~.\eeqa

\noindent When $a=0$, one recovers the standard expression

\beq {\nu' \over 2} = G{ m_e + 4 \pi p r^3 \over r^2 - 2Gm_e r}~~.\eeq

\noindent Linearizing with respect to $a$ yields 

\beq {\nu' \over 2} = G { m_e + 4 \pi p r^3 \over r^2-2Gm_e r} + a h(p,\rho)~~, \eeq

\noindent where the function $h(p,\rho)$ is defined through

\beqa && h(p, \rho) = \\* \nonumber && \left( {5 \over 8}p'' - 4 \pi G p \rho \right) r + \left( {5 \pi \over 2} G \rho r^2 -{3 \over 8}{G m_e \over r} - {1 \over 4} \right) p' \approx \\* \nonumber && \left( {5 \over 8}p'' - 4 \pi G p \rho \right) r -{p' \over 4}~~, \eeqa

\noindent after considering the inequality $r \gg 2Gm_e$ and also $5\pi G\rho r^2 /2 \ll 1$; taking the Sun's maximum, central, density $ \rho_c = 1.622 \times 10^5~kg/m^3$, and the radius $R_\odot = 6.955 \times 10^8~m$, one gets $5\pi G\rho r^2 /2 c^2 = 4.57 \times 10^{-4} \ll 1$.

Before continuing, one can rewrite the expression for the gravitational mass, obtained before, but imposing the limit $e^{-\si} \approx e^{-\nu} \approx 1$, $\nu' \approx \si' \approx 0$, which yields

\beq \square_{00} p \approx 0~~~~,~~~~ \square_{rr} p \approx p'' ~~~~,~~~~ \square_{\th\th} p \approx r p'~~. \eeq

\noindent This approximation, together with $p \ll \rho$, implies that

\beq m'_e \approx 4 \pi r^2 \rho + ar^2 \left( {3 p''\over 8G} +{p' \over 4 Gr} - 2 \pi \rho^2 \right)~~.\eeq

Similarly, one obtains

\beqa R &\approx& {-\rho - 3 a \ka \left( p'' + {2 \over r} p' \right) \over \ka \left( 2 + a \rho \right) } \approx \\* \nonumber && - {\rho \over 2 \ka } - {a \over 2 } \left[ 3 \left( p'' + {2 \over r} p' \right) - {\rho^2 \over 2 \ka} \right]~~. \eeqa

Although the expression for the scalar curvature $R$ is somewhat involved, one can approximate its derivative by the unperturbed value $R' = - \rho' / 2 \ka$, losing only terms of order $O(a^2)$; indeed, one could write 

\beqa R' &=& -{\rho' \over 2 \ka} + a F(\rho,p) \rightarrow \\* \nonumber \la R' & = & -{\la \rho' \over 2 \ka} + a \la F(\rho,p) = - {a \over 2 } \rho' + a^2 {F(\rho,p) \over \ka} ~~, \eeqa

\noindent with 

\beq F(\rho,p)= - {1 \over 2 } \left[ 3 \left( p'' + {2 \over r} p' \right) - {\rho^2 \over 2 \ka} \right]'~~.\eeq

To write the modified equation of hydrostatic equilibrium, one now resorts to the non-(covariant )conservation of the energy-momentum tensor, Eq. (\ref{non-cons}). The expression for the covariant derivative is purely geometric; in order to derive an expression for $p' $, one aims at the $\nu = r$ component of the above equation; with our choice of $f_2 = R$ and $\mathcal{L}_m = p$, one gets the modified TOV equation

\beqa && p' + {\nu' \over 2} (\rho + p) = {2 \la \over 1 + \la R }\left(1 - {2Gm_e \over r} \right) R' \rightarrow \\* \nonumber && -{\nu' \over 2}(\rho + p) \approx p'- {2 \la p \over 1 + \la R } R' \approx p' + a p \rho' ~~, \eeqa

\noindent dropping higher order terms and using $R' = - \rho' / 2 \ka$.

Thus, one finally obtains a set of three differential equations for the problem at hand:

\beqa \label{eqset1} p' &-& 2 \la p R' = -{\nu' \over 2} (\rho + p)~~, \\* \label{eqset2} m'_e &=& 4 \pi r^2 \rho + ar^2 \left( {3 p''\over 8G} +{p' \over 4 Gr} - 2 \pi \rho^2 \right)~~, \\* \label{eqset3} {\nu' \over 2} &=& G{ m_e + 4 \pi p r^3 \over r^2 -2 Gm_er} + a \left[ \left( {5 \over 8}p'' - 4 \pi G p \rho \right) r - {p' \over 4} \right]~~. \eeqa 

\noindent Substituting Eq. (\ref{eqset1}) into (\ref{eqset3}), after some algebra one gets the modified TOV equation,

\beqa \label{TOVeq} &&p' + G (\rho + p){ m_e +4 \pi p r^3 \over r^2 - 2Gm_er } = \\* \nonumber && a \left[ \left( \left[ {5 \over 8} p'' - 4 \pi G p \rho \right] r - {p' \over 4} \right)\rho+ p \rho' \right] ~~. \eeqa

\noindent This yields the non-relativistic hydrostatic equilibrium equation,

\beq \label{hydroeq} p' + {Gm_e \rho\over r^2 } = a \left[ \left( \left[ {5 \over 8} p'' - 4 \pi G p \rho \right] r - {p' \over 4} \right) \rho+ p \rho' \right] ~~. \eeq

\subsection{Polytropic equation of state}

Realistic stellar models rely on four differential equations, together with appropriate definitions \cite{books}; aside from the mass conservation condition (\ref{eqset2}) and the hydrostatic equilibrium equation (\ref{TOVeq}) or (\ref{hydroeq}), these express energy conservation and transport, through

\beq {d L(r) \over dr} = 4 \pi r^2 \ep \rho(r) \eeq

\noindent and

\beq {d T(r) \over dr} = - {L(r) \over 4\pi r^2 \la_c}~~, \eeq

\noindent respectively. In the above, $L(r)$ is the energy flow across a sphere of radius $r$, $\ep$ is the energy generation rate per mass unit, and $\la_c$ is the conductivity coefficient. For given $\ep$ and $\la_c$, which account for the processes ongoing inside the star, one is left with five unknowns, $m'_e$, $p$, $\rho$, $T$ and $L$; an additional relation is needed, in the form of a suitable equation of state $p = p (\rho)$. 

Many candidate equations of state and solar models are available, with varying degrees of sofistication, accounting for effects such as chemical composition, solar matter mixing, discontinuities between layers, heat difusion, {\it etc.}; two outstanding examples are the Mihalas-Hummer-Dappen and OPAL equations of state \cite{EOS}. However, solving the above set of differential equations with a realistic equation of state requires heavy-duty numerical integration with complex code designs, and an analysis of the perturbations induced by a non-minimum coupling between matter and curvature is beyond the scope of the present study.

Instead, one may resort to a very simplistic assumption, the so-called polytropic equation of state. This is commonly given by $p = K \rho_B^{(n+1)/n}$, where $K$ is the polytropic constant, $\rho_0$ is the baryonic mass density and $n$ is the polytropic index \cite{books,Tooper1,Stergioulas}. The polytropic index $n$ interpolates between the basic thermodynamical processes: $n = -1$ for isobaric, $n = 0$ for isometric, and an infinite polytropic index $n$ for an isothermal sphere. Adiabatic processes yield $n = 1/(\ga_A -1)$, with $\ga_A = c_p/ c_V$ the adiabatic coefficient. Several crude approximations to relevant astrophysical systems are also obtained: $n = 3/2$ may model the degenerate star cores found in giant (gaseous) planets, white or brown dwarfs,and red giants; a polytropic index $n = 5$ yields a boundless system (that is, with non-vanishing density everywhere), which was taken by Schuster as the first candidate for a stellar system. Finally, a polytropic equation of state was used by Eddington in his proposal for the first solar model, with $n = 3$; naturally, it does not offer an accurate description of the solar interior, and has been deprecated by the following advancements.

Nonetheless, the use of a polytropic equation of state is still of interest, due to its simplicity and ease of manipulation, which render it a valuable tool in more theoretically driven studies, as is the present case (as an example, the generalized Chaplygin gas is a polytrope with index $n =-1/(1+\al)$ \cite{darkstar}). Clearly, more realistic assumptions regarding the structure of the Sun would improve the final results; also, the procedure outlined in this work could also be applied to more exotic bodies, either through the use of an adequate polytropic index, or via a more realistic equation of state, possibly yielding a more stringent constraint on the coupling parameter $\la$.

Recall that the effective mass $m_e$ is defined in terms of the energy density $\rho$, which appears in the energy-momentum tensor; these two quantities are related through $\rho = \rho_B + n p$, yielding the relation 

\beq \rho = \left({p \over K}\right)^{n / ( n+1 )} + n p~~. \eeq 

\noindent However, since one is interested in probing the Newtonian regime which occurs in the Sun, the condition $p \ll \rho$ holds; therefore, $\rho \simeq \rho_B$, and one may take the form $p = K \rho^{(n+1)/n}$ for the equation of state \cite{Tooper2}. 

In order to transform the modified hydrostatic equilibrium Eq. (\ref{hydroeq}) into a differential equation with a single variable, one first rewrites Eq. (\ref{hydroeq}) as
 
\beqa \nonumber && {1 \over r^2}\left[ {r^2 \over \rho } \left( p' + a \left[ \left( \left[ {5 \over 8} p'' - 4 \pi G p \rho \right] r - {p' \over 4} \right) \rho+ p \rho' \right] \right) \right]' \\* && = -4 \pi G \rho - a \left( {3 p''\over 8} +{p' \over 4 r} - 2 \pi G \rho^2 \right) ~~, \eeqa 

\noindent and inserts the polytropic equation of state, $p = K \rho^{(n+1)/n}$, written as $\rho = \rho_c \th^n (\xi)$ and $p = p_c \th^{n+1} ( \xi) $, with $\xi = r / r_0$ a dimensionless variable and $r_0^2 \equiv {(n+1)} p_c / 4 \pi G\rho_c^2 $; as stated before, $ \rho_c = 1.622 \times 10^5~kg/m^3$ is the central density, and $p_c = 2.48 \times 10^{16}~Pa$ is the central pressure. One obtains the perturbed Lane-Emden equation for the function $\th(\xi)$:

\begin{widetext} \beq \label{LE} {1\over \xi^2} \left[ \xi^2 \th' \left( 1 + A_c \th^n \left[ \left[{5 \over 8} \left( \th'' + n {\th'^2 \over \th} \right) - N_c \th^{n+1} \right] { \xi \over \th' } + {3n-1 \over 4(n+1) } \right]\right) \right]' = - \th^n \left[1 + A_c \left( {3 \over 8}\left[ \th'' + n {\th'^2 \over \th} \right] + { \th' \over 4 \xi} - {\th^{n } \over 2} \right) \right] ~~,\eeq
\end{widetext}

\noindent where the prime now denotes derivation with respect to the dimensionless radial coordinate $\xi$, and one has defined $A_c \equiv a \rho_c$ and $N_c \equiv p_c / \rho_c = 1.7 \times 10^{-6}$, for convenience. Obviously, setting $A_c = 0$ one recovers the unperturbed Lane-Emden equation \cite{books}

\beq \label{LE0} {1 \over \xi^2} \left(\xi^2 \th' \right)' = - \th^n~~.\eeq 

One may try to solve Eq. (\ref{LE}) analytically around $\xi = 0$, and compare with the solution of the unperturbed equation, given (in the vicinity of $\xi = 0$) by \cite{books}

\beq \th(\xi) \approx 1 - {1 \over 6} \xi^2 + {n \over 120} \xi^4~~. \label{eq0} \eeq

\noindent This calculation (outlined in Appendix B) yields  

\beq \th(\xi) \approx 1 -A \xi^2 +B\xi^4 ~~, \label{origin} \eeq

\noindent with

\beq A = { 1 \over 6} \left( 1 - A_c { 13 + 25 n \over 12(n+1) } \right)~~,\eeq

\noindent and
 
\beq B= {n \over 120} \left ( 1 - A_c {11 \over 60} {39 + 59n \over n+1} \right) ~~,\eeq

\noindent neglecting the $N_c \ll 1$ term and taking the limit $|A_c| \ll 1$. Taking $A_c=0$ gives back the unperturbed values $A= 1/6$ and $B=n/120$, as expected.

For $n = 0$ (a constant density model), one finds

\beq \th(\xi) \approx 1- { 1 \over 6} \left( 1 - { 11 \over 6 } A_c \right) \xi^2 ~~. \eeq

\noindent For $n=3$ (the first proposed model for the Sun), one obtains

\beq \th(\xi) \approx 1- { 1 \over 6} \left( 1 - { 11 \over 6 } A_c \right) \xi^2 + {1 \over 40} \left ( 1 - {99\over10} A_c \right) \xi^4 ~~. \eeq

\noindent Finally, for $n \rightarrow \infty$ (an isothermal sphere), one gets

\beq \th(\xi) \approx 1- { 1 \over 6} \left( 1 - { 23 \over 12 } A_c \right) \xi^2 + {1 \over 40} \left ( 1 - {1837\over180} A_c \right) \xi^4 ~~. \eeq

\subsection{Perturbative solution}

Inspection of Eq. (\ref{LE}) shows that it is a third-order equation, while the unperturbed Lane-Emden equation is only second-order. Thus, the initial conditions $\th(0) = 1$ and $\th'(0) = 0$ alone are not sufficient. Indeed, an extra condition regarding the initial behaviour of the second derivative must be provided; this is derived from the series expansion of $\th$ around $\xi = 0$ taken before, that is

\beq \th''(0) = -{1 \over 3}\left(1 - A_c {13 + 25n \over 12(n+1)}\right) ~~. \eeq

In order to numerically solve the perturbed Lane-Emden equation, one rewrites Eq. (\ref{LE}) (after a long derivation) as

\beqa && \th''' = - {8 \over 5 A_c \xi} \left[ {\th'' + {2 \over \xi} \th' \over \th^n} + 1 \right] - n (n - 1) {\th'^3 \over \th^2} \\* \nonumber && + {24 \over 5} N_c { \th^{n+1} \over \xi} - {2 \over 5} \left[ 4 {3n +2 \over n+1} {\th'' \over \xi} + {7n -1 \over n+1} {\th' \over \xi^2} \right] + \\* \nonumber && {4 \over 5} \th^n\left[ {1 \over \xi} + 2 (2n+1) N_c\th' \right] - n { \th' \over \th } \left[ 3 \th'' + {8 \over 5} {3n+2 \over n+1} {\th' \over \xi} \right] \\* \nonumber && ~~. \eeqa 

\noindent Clearly, when $A_c \rightarrow 0 $, the first term blows up unless it is compensated by the condition $\th'' + 2 \th' / \xi = - \th^n$, which is precisely the unperturbed Lane-Emden equation. By the same token, the first term expresses the deviation from the unperturbed case, and should be of order $A_c$, therefore cancelling out the divergence.

However, implementing the above third order differential equation proves too computationally demanding; thus, one must first approach the perturbed Lane-Emden equation and, given that one is searching for a perturbation, expand it in terms of $\th = \th_0 (1 + A_c \de)$, where $\th_0$ is the solution to the unperturbed Lane-Emden equation and $\de$ is the (relative) perturbation, obeying $|A_c| \de \ll 1$. Thus, one may write, for $n > 0$,

\beq \th^n = \th_0^n (1 + A_c \de)^n \approx \th_0^n \left( 1 + nA_c \de \right) ~~. \eeq

\noindent If this expansion is introduced in the perturbed Lane-Emden equation, and considering only terms of order $A_c$, one obtains a differential equation for $\de$, with $\th_0$ as source:

\beqa && \de'' + 2 \left( {\th_0' \over \th_0} + {1 \over \xi} \right) \de' + (n-1) \th_0^{n-1} \de = {5n\over 2} \xi \th_0^{2 n-2} \th_0' \\* \nonumber && + (2 n +1) N_c \xi \th_0^{2 n-1} \th_0' + {9 n+5 \over 4 (n+1)} \th_0^{2 n-1} + 3 N_c \th_0^{2n} \\* \nonumber && - {5n (n-1) \over 8} \xi \th_0^{n-3} \th_0'^3+ {n(3n+7) \over 4(n+1)} \th_0^{n-2} \th_0'^2 + {1 \over 2}{\th_0^{n-1} \th_0' \over \xi} ~~.\eeqa

\noindent having eliminated the second derivative of the unperturbed solution through Eq. (\ref{LE0}). From Eq. (\ref{origin}), one concludes that this differential equation is supplemented by the initial conditions $\de(0)=0$ and $\de'(0)=0$.

Note that $\de$ does not depend on $A_c$: one must only find the unique $\de$ (for each $N_c$ and $n$); one aims at a simultaneous variation of the model's parameter $A_c$ and the polytropic index $n$ in the vicinity of the standard solar value $n=3$, enabling the plotting of an exclusion graph in the $(A_c, n)$ plane. For $n=3$, the differential equation for $\de$ simplifies to

\beqa && \de'' + 2 \left( {\th_0' \over \th_0} + {1 \over \xi} \right) \de' + 2 \th_0^2 \de = 3 N_c \th_0^6 + 2 \th_0^5 + \\* \nonumber && 7 N_c \xi \th_0^5 \th_0'+ {15\over 2} \xi \th_0^4 \th_0' + {1 \over 2\xi}\th_0^2 \th_0' + 3 \th_0 \th_0'^2 - {15 \over 4} \xi \th_0'^3 ~~.\eeqa

\subsection{Mass budget and matching conditions}

One now computes the deviation between the effective mass $m_e$, defined in Eq. (\ref{mass'}), and the gravitational mass, defined by $m'_g = 4 \pi r^2 \rho$; some algebra yields

\beqa \label{massdif} m'_e - m_g' & = &{n+1 \over G} A_c N_c \xi^2 \th_0^n \left[- {3 n \over 8} {\th_0'^2 \over \th_0} + {\th_0' \over 2 \xi} + {7 \over 8} \th_0^n \right] ~~, \eeqa

\noindent which is explicitly written in terms of $A_c = a \rho_c$, the model's dimensionless parameter, and $N_c \equiv p_c / \rho_c$, the ratio that measures the validity of the Newtonian regime. One has also to eliminate $\de''_0$ through Eq. (\ref{LE0}); one concludes that not only is the above mass difference small (due to the perturbative regime, $|A_c| \ll 1$), but that it is further suppressed by the factor $N_c \ll 1$.

Clearly, the model's parameters should be adjusted to the known observables: the star's radius $R_\odot$ and mass $M$. The issue of this identification is rather delicate; indeed, $M$ should not be identified with the gravitational mass $m_g(R_\odot)$, but with the total effective mass $m_e(R_\odot)$. Asides from the physical interpretation of the full energy content of the star affecting geodetic motion around it, one can also resort to the matching conditions of the inner metric with the outer Schwarzschild metric; indeed, writing $M_e \equiv m_e(R_\odot)$, $M_g \equiv m_g(R_\odot)$, $M'_e \equiv m'_e(R_\odot)$ and $M'_g \equiv m'_g(R_\odot)$, one has

\beqa \si_{-}' (R_\odot)& = & -2G {M_e - M'_e R_\odot \over R_\odot(R_\odot-2GM_e)}~~, \\* \nonumber \si_{+}' & = & - 2{GM \over R_\odot(R_\odot-2GM) }~~, \\* \nonumber \nu_{-}' (R_\odot) & = & 2{G M_e \over R_\odot(R_\odot-2GM_e)} + \\* \nonumber && {a \over 4} \left({5 \over 2} p''(R_\odot) R_\odot - p'(R_\odot) \right)~~, \\* \nonumber \nu_{+}' & = & - \si_{+}' = 2{GM \over R_\odot(R_\odot-2GM)} ~~, \eeqa

\noindent where the $+$ and $-$ subscripts indicate inner or outer boundary condition. If one identifies $M = M_e$, it follows that $M'_e=0$, which implies

\beq {3 p'' (R_\odot)\over 2} +{p' (R_\odot)\over R_\odot} = 0~~, \eeq

\noindent (after resorting to Eq. (\ref{mass'}), with $\rho(R_\odot) = 0$), and

\beq {5 \over 2}p''(R_\odot) R_\odot -p'(R_\odot) = 0 ~~. \eeq

\noindent Clearly, both conditions hold, since

\beqa p'(R_\odot) &\propto& \th'(\xi_1) \th^n(\xi_1) = 0~~, \\* \nonumber p'' (R_\odot)&\propto& \th''(\xi_1)\th(\xi_1)^n + n\th'(\xi_1)^2 \th(\xi_1)^{n-1} =0~~, \eeqa

\noindent where $\xi_1$ signals the star's boundary, through $\th(\xi_1) = 0$.

\subsection{A solution for the divergence problem}

Given the identification $M=M_e \equiv m_e(R_\odot)$, one can now deduce the perturbation to the central density $\rho_c$, which is a model dependent parameter. However, before using Eq. (\ref{LE}) to evaluate the perturbation $\de$ and extract relevant quantities, it should be noticed that, by inspection, it is clear that $\de$ diverges when $\th_0$ approaches zero. Hence, one cannot extend the perturbational approach to the full range of the star, and should instead deal with the full differential equation for the perturbed $\th$.

In order to circumvent this issue, recall that there is a pronounced deviation between the predictions of the polytropic model and the realistic Standard Solar Model, for $r > R_r = 0.713~ R_\odot$; this reflects the crossing from the radiative zone, where the $n=3$ polytrope is a good approximation for the Sun, and the convection zone, where this approach fails. Hence, the irregular behaviour of $\de$ near $\xi_1$ may be safely disregarded, since the fundamental equation for $\th$ is not valid there. Instead, one shall consider only the range $ 0 \leq r \leq R_r $ or, equivalently, $0 \leq \xi \leq \xi_r = 0.713~ \xi_1$. In doing so, one is of course neglecting the contribution of the latter for all relevant quantities: however, although the density and pressure are still significant at that point, both the polytropic and the Standard Solar Model show that up to $ 99.1 \% $ of the Sun's mass is located within the radiative zone.

Also,it can be numerically shown that the boundary condition $\th(\xi) = \th_0(\xi)(1+A_c\de(\xi))=0$ does not shift significantly from the unperturbed $\th_0(\xi_1)=0$ case, so that no problem arises from neglecting any changes to the scale factor $r_0$ (which, recall, relates the dimensionless coordinate $\xi$ with the physical distance to the center $r$). Furthermore, the matching of the inner and outer derivatives of the metric should not be taken as a realistic condition, but merely a consistency check for the developed model.

\subsection{Model-dependent parameters}

One proceeds with the calculation of the total mass of the Sun or, more accurately, the mass of the radiative zone: for the unperturbed case, one has (see Appendix C for the derivation of the following results)

\beq M = -4 \pi {R_r^3 \over \xi_r} \rho_{c0} \th_{0r}' \rightarrow \rho_{c0} = -{M \over 4 \pi \th_{0r}' } {\xi_r \over R_r^3 }~~. \eeq

\noindent where one defines $\th'_{0r} \equiv \th_0'(\xi_r)$ and $\rho_{c0}$ is the unperturbed central density (that is, obtained from $M$, $R$ and the numerical results for the unperturbed solution $\th_0$).

In the perturbed case, one obtains

\beqa && M = -4 \pi {R_r^3 \over \xi_r} \rho_c \th_{0r}' \times \\* \nonumber && \left[ 1 - {A_c \over \xi_r^2 \th_{0r}' } \int_0^{\xi_r} \xi^2 \th_0^n \left(n \de + {3 n \over 8} {\th_0'^2 \over \th_0} - {\th_0' \over 2 \xi} - {7 \over 8} \th_0^n \right) ~d\xi \right] ~~. \eeqa

\noindent Notice that there are two $A_c$-dependent contributions: one arising from $\th_0$ and its derivatives, since including the perturbation $\de$ would only amount to second-order terms $O(A_c^2)$, and other involving the integral of $\de$. 

Since the total mass $M$ does not change (only its interpretation as a purely gravitational mass or a sum of gravitational plus ``active'' components), one gets

\beqa && 1 - {\rho_{c0} \over \rho_c} = \\* \nonumber && {A_c \over \xi_r^2 \th_{0r}' } \int_0^{\xi_r} \xi^2 \th_0^n \left[n \de + {3 n \over 8} {\th_0'^2 \over \th_0} - {\th_0' \over 2 \xi} - {7 \over 8} \th_0^n \right] ~d\xi~~. \eeqa

Clearly, taking $A_c =0$ yields $\rho_c = \rho_{c0}$. Also, notice that $\xi_r$ is not perturbed: this would reflect a change in $\xi_1$ (since $\xi_r = 0.713~ \xi_1$), which has already been ruled out. Also, notice that the denominator affecting the integral is equal to ${ \xi_r^2 \th_{0r}' = \int_0^{\xi_r} (\xi^2 \th_0')' = -\int_0^{\xi_r} \th_0^n}$: thus, one can interpret this integral as the volume averaging of the expression in brackets within the integrand, with the density as weighting function, that is, $\th^n \propto \rho$.

From the polytropic equation of state, one gets $ \rho \propto T^{n+1} $, which enables writing

\beqa && 1 - \left({T_{c0} \over T_c}\right)^{n+1} = \\* \nonumber && {A_c \over \xi_r^2 \th_{0r}' } \int_0^{\xi_r} \xi^2 \th_0^n \left[n \de + {3 n \over 8} {\th_0'^2 \over \th_0} - {\th_0' \over 2 \xi} - {7 \over 8} \th_0^n \right] ~d\xi ~~. \eeqa

\section{Numerical results}

\begin{figure}
\epsfysize=5.7cm
\epsffile{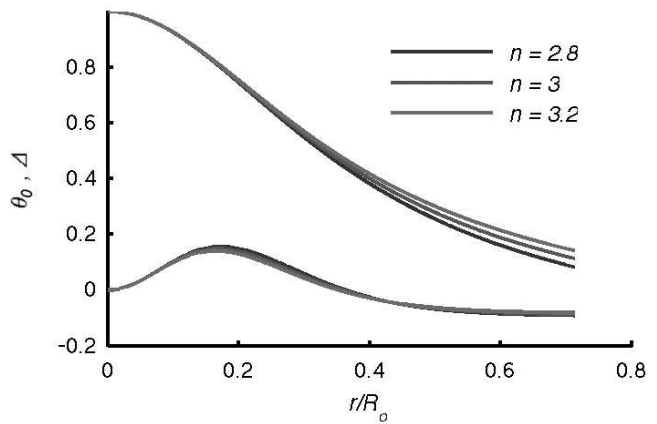}
\caption{Profiles for the unperturbed solution $\th_0$ (top) and absolute perturbation $\De = \de \th_0 $ (bottom), for $ 2.8 \leq n \leq 3.2 $. }
\label{absprofiles}
\end{figure}

\begin{figure}
\epsfysize=5.7cm
\epsffile{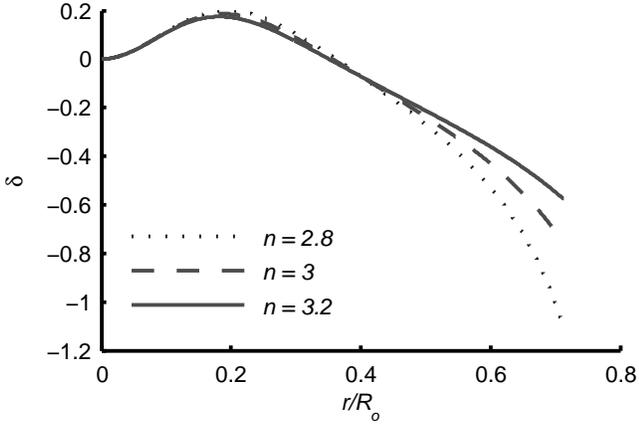}
\caption{Profile of the relative perturbation $\de$ for $ 2.8 \leq n \leq 3.2 $. }
\label{relprofiles}
\end{figure}

\begin{figure}
\epsfysize=5.7cm
\epsffile{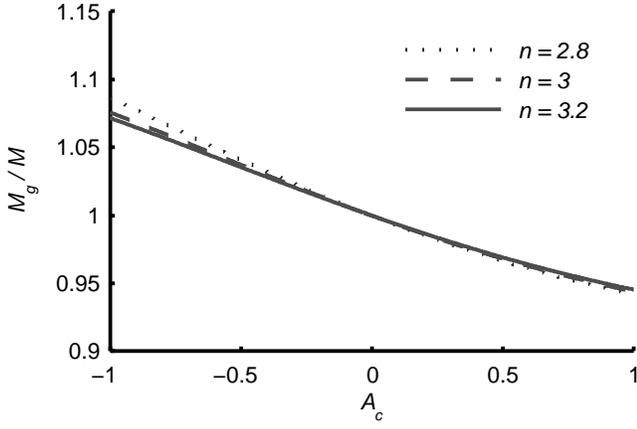}
\caption{Fraction of gravitational to total mass as a function of $A_c$, for $ 2.8 \leq n \leq 3.2 $.}
\label{gravmass}
\end{figure}

\begin{figure}
\epsfysize=5.7cm
\epsffile{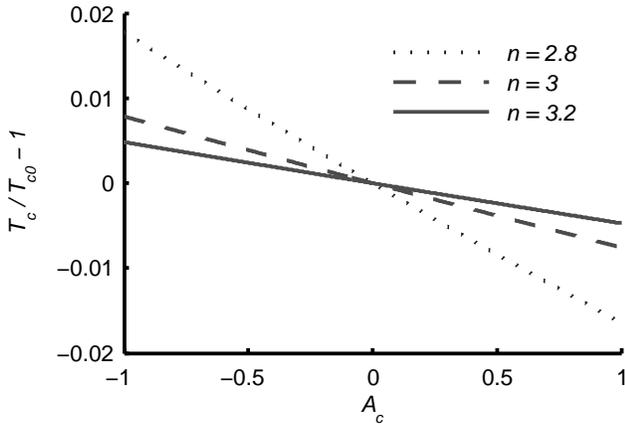}
\caption{Relative deviation of the central temperature $T_c / T_{c0}-1$ as a function of $A_c$, for $ 2.8 \leq n \leq 3.2 $.}
\label{relTc}
\end{figure}

\begin{figure}
\epsfysize=5.7cm
\epsffile{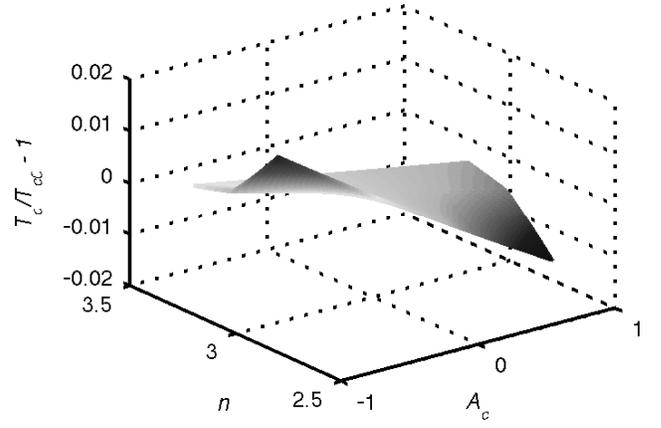}
\caption{Relative deviation of the central temperature $T_c / T_{c0}-1$ as a function of $A_c$ and $n$.}
\label{relTc3d}
\end{figure}

\begin{figure}
\epsfysize=5.7cm
\epsffile{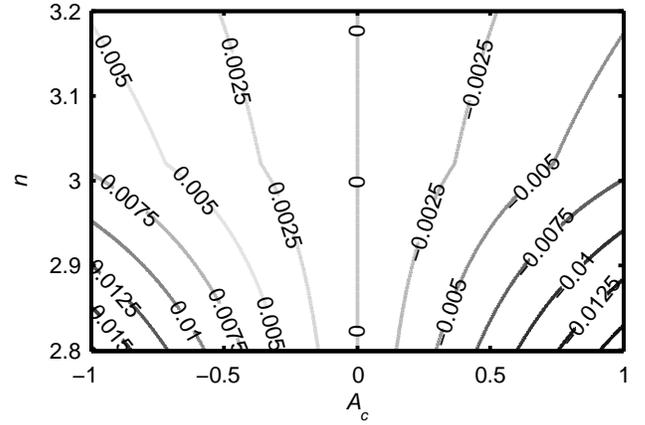}
\caption{Contour plot for the relative deviation of the central temperature $T_c / T_{c0}-1$ as a function of $A_c$ and $n$, with contour lines of step $0.1 \% $.}
\label{relTc2d}
\end{figure}

\begin{figure}
\epsfysize=5.7cm
\epsffile{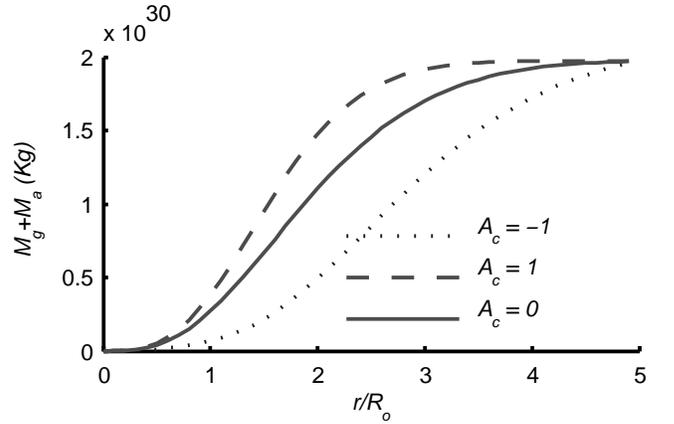}
\caption{Mass profiles, for $n=3$ and $A_c = -1,~0,~1$.}
\label{mass}
\end{figure}

\noindent The above results allow us to construct an exclusion plot for the central temperature on the $(n, A_c)$ parameter space, by imposing the constraint $1 - T_c/T_{c0} \leq 0.06~(6 \%)$, the uncertainty of the central temperature of the Sun. Solving numerically the differential equation for $\de$ for a varying $n$, and computing all relevant quantities for varying $A_c$, and defining the absolute perturbation $\De \equiv \th_0 \de$, so that $\th = \th_0(1+A_c \de) = \th_0 + A_c \De$, one obtains the results shown in the Figures \ref{absprofiles} to \ref{mass}. 
 
First, notice that, from Fig. \ref{absprofiles}, the absolute perturbation $\De$ is fairly insensitive to the value of $n$; clearly, given that the profile of $\th_0$ for varying $n$ differs more sharply as $\xi \rightarrow \xi_{r}$ (as seen in Fig. \ref{absprofiles}), a approximately constant $\De$ translates into a relative perturbation $\de$ that also exhibits this behaviour, as can be seen from Fig. \ref{relprofiles}. Furthermore, notice the indication of the divergence of the relative perturbation $\de$ in Fig. \ref{relprofiles} (which does not depend on $n$, since $\th_0 \rightarrow 0$ and $\De \rightarrow ~const.$ yields $\de \rightarrow \infty$, as previously discussed); as stated before, this divergence is avoided by dealing only with the radiative region $r \leq R_{r}$. 

Also, as can be seen from Fig. \ref{relprofiles}, the relative perturbation $\de$ peaks at $|\de|_{max} \approx 1.1$ (for $n = 2.9$, and a smaller value of $|\de|_{max} \approx 0.8$ for $n=3$); hence, the perturbative condition $|A_c| \de \ll 1 $ translates to $|A_c| \ll 1$, leading to the chosen (extreme) interval for simulation $-1 \leq A_c \leq 1$. Hence, according to Figs. \ref{relTc}, \ref{relTc3d} and \ref{relTc2d}, no relative deviation of the central temperature above the experimentally determined level of $6\%$ is attained. However, the values found, of the order of $1\%$, indicate that any future refinement of the experimental error of $T_c$ could yield a direct bound on the parameter $A_c$.

The above results show that $A_c$ is presently unconstrained by solar observables, aside from the perturbative condition $|A_c| \ll 1 \rightarrow |\la| \ll \ka /\rho_c$; taking $\ka = c^4/16 \pi G = 2.41 \times 10^{42}~kg.m/s^2 $ and $\rho_c = 1.622 \times 10^5~kg/m^3$, this yields $|\la| \ll 1.48 \times 10^{37}~ m^4/s^2$ or, in natural units, $|\la| \ll 4.24 \times 10^{33}~eV^{-2} $. However, this study assumes that the effects arising from the non-trivial matter-geometry coupling supersede those from the non-linear curvature term in the modified Hilbert-Einstein action (\ref{action}), as expressed by the inequalities (\ref{ineqa}), (\ref{ineqf1a}) and (\ref{ineqf1b}). Having numerically determined the perturbative solution $\de$, one can now re-evaluate these conditions. For this, one first replaces $R$ by its unperturbed value (with the Newtonian approach $\rho \gg p$), $R \approx -\rho/2\ka $, since corrections would amount to second-order terms in $\la$.

Considering the scaling laws, $\rho = \rho_c \th^n (\xi)$ and $p = p_c \th^{n+1} ( \xi) $, and the inequalities (\ref{ineqf1a}) and (\ref{ineqf1b}), some algebra (see Appendix D) yields, for the $f_1 = f_{1a} \equiv 2 \ka( R - \al R^{1-m})$ case,

\beqa \label{ineqla1} |\la| & \gg & \left({2 \ka \over \rho_c \th^n }\right)^{m+1} {\al \over 2 N_c \th} ~~, \\* \nonumber |\la| & \gg & { n \over n+1} \left({2 \ka \over \rho_c \th^n} \right)^{m+1} {\al (1-m) m \over 2 N_c \th }~~, \\* \nonumber |\la | & \gg & {n \over n+1} \left({2 \ka \over \rho_c \th^n} \right)^{m+1} { \al (1-m)m \over 2 N_c \th } C_{n,m} (\xi) ~~,\eeqa

\noindent while, for the $f_1=f_{1b} \equiv 2\ka( R + \al R^2)$ case,

\beqa \label{ineqla2} \la & \gg & {\al \over 2 N_c \th} ~~, \\* \nonumber \la& \gg & {n \over n+1} {\al \over N_c \th} ~~, \\* \nonumber \la & \gg & {n \over n+1} {\al \over N_c \th} C_{n,-1}(\xi) ~~,\eeqa

\noindent where one defines the quantity

\beq C_{n,m} (\xi) \equiv \left| {(1+ nm) \th'^2(\xi) - \th \th''(\xi) \over n \th'^2(\xi)+ \th \th''(\xi)} \right| ~~. \eeq

One can plot the function $C_{n,m}(\xi)$ by evaluating $\th$ by the unperturbed solution $\th_0$, using Eq. (\ref{LE0}) and taking $n \approx 3$. One obtains

\beq C_{n,m} (\xi) \approx \left| { \th_0' \left[(1+ 3m)\th_0' + {2\over \xi}\th_0 \right] + \th_0^4 \over \th_0' \left(3\th_0' + {2 \over \xi} \th_0 \right) + \th_0^4}\right| ~~. \eeq

\begin{figure}
\epsfxsize=9cm
\epsffile{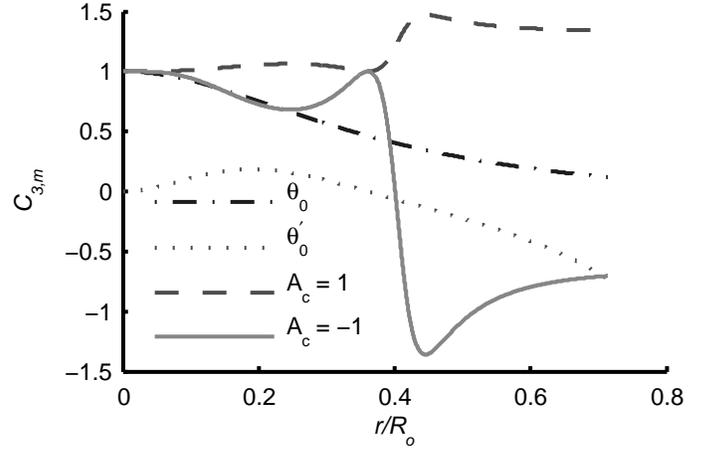}
\caption{Profiles of $C_{3,m}$, for $m = -1,~0,~1$, superimposed on profile of $\th_0$ and $\th_0'$, with $n=3$.}
\label{Cnmgraph}
\end{figure}

\noindent Varying $m$ in its domain $0<m<1$ (for the $f_1=f_{1a}$ case) and $m=-1$ (for the $f_1= f_{1b}$ case); its profile is depicted in Figure \ref{Cnmgraph}. One concludes that $|C_{n,m} | \lesssim 1.5$, for both the $f_1 = f_{1a} $ case ($0<m<1$) as well as the $f_1 = f_{1b}$ case ($m=-1$). This, together with the constraint $\th(\xi) \gtrsim 0.1$ (as can be seen from Fig. \ref{absprofiles}) and the quantity $N_c \equiv p_c / (\rho_c c^2)= 1.7 \times 10^{-6}$ yields, for the $f_1 = f_{1a}$ case,

\beq \label{ineqla3} |\la| \gg 2.8 \times 10^{43} (8.48 \times 10^{36})^m (1-m)m \al ~eV^{-2(1+m)} ~~,\eeq

\noindent while, for the $f_1=f_{1b}$ case,

\beq \label{ineqla4} |\la| \gg 6.6 \times 10^6 \al~~. \eeq

\noindent Using the previously discussed value, $\al=(10^{12}~GeV)^{-2}$ , which arises from the Planck-scale renormalization of the theory, one gets

\beq \label{ineqla5} |\la| \gg 6.6 \times 10^{-36}~eV^{-2} = (3.9 \times 10^8~GeV)^{-2}~~. \eeq

Also, a recent paper has reported a relation between the coefficient of the quadratic term in $f_1$ and the PPN parameter $\ga$ \cite{Capozziello},

\beq \al = {1 \over 2 \ka} \sqrt{\left| {1 - \ga \over 2\ga -1} \right|} ~~. \eeq

\noindent Hence, the current experimental constraint $\ga - 1 = (2.1 \pm 2.3)\times 10^{-5}$ \cite{Cassini} yields 

\beq \al \leq 0.17 M_{Pl}^{-2} = \left( 3.0 \times 10^{19}~GeV \right)^{-2}~~, \eeq

\noindent where $M_{Pl}= \sqrt{\hbar c /G }= 1.2 \times 10^{19} ~GeV$ is the Planck mass. This yields 

\beq \label{ineqla6} |\la| \gg 7.4 \times 10^{-51}~eV^{-2} = (1.2 \times 10^{16}~GeV)^{-2}~~. \eeq

Clearly, the $f_1=f_{1b}$ case does not impose any significant bound on $\la$, with the above value laying many orders of magnitude below the upper bound, $|\la| \ll 4.24 \times 10^{33}~eV^{-2} $, arising from the perturbative treatment.

Notice that the above inequalities (\ref{ineqla1}), (\ref{ineqla2}), (\ref{ineqla3}), (\ref{ineqla4}), (\ref{ineqla5}) and (\ref{ineqla6}) are not to be considered as restrictions on the parameters, but as conditions for the validity of the regime where the effects of the matter-geometry coupling supersede those of the the non-linear curvature term in Eq. (\ref{action}).

\section{Discussion}

In this work we have examined a model where the usual Einstein-Hilbert action is modified, not only by allowing non-linear curvature terms to appear, but by also enabling an explicit, non-minimum coupling between the curvature and the Lagrangian of matter fields. We first assume that this matter-geometry coupling is linear in curvature and its effect allows the non-linear curvature terms to be neglected -- an assertion that is qualified in the end of the study, for two relevant non-linear $f(R)$ models. 

We then proceed and perform the necessary calculations to ascertain the effect of the latter in the hydrostatic equilibrium of an $n \approx 3$ polytrope such as the Sun. We assume a perturbative regime to the usual Tolman-Oppenheimmer-Volkoff equation, and take into consideration the validity of the Newtonian regime in this modified theory, as well as the redefinition of relevant quantities, which are computed numerically and compared with Solar observables. This goal is achieved through the use of the (non-relativistic) polytropic equation of state $p = K \rho_B^{(n+1)/n}$: as stated before, this is a very simplistic description of the complex behaviour of solar matter, and has been superseded by much more elaborate equations of state; we use it in our study so to better illustrate the effects of the matter-curvature coupling on an easily understandable, closed model. However, we should remark that this restrictive treatment, although advantageous from the theoretical point of view, might conceal some of the more intricate phenomenology found in stellar systems, which could affect our results. Clearly, a subsequent study should consider a more realistic solar model.

The results allow us to conclude that no strong constraints on the matter-geometry coupling from the comparison between the model's predictions and current experimental sensitivity, aside from the perturbative approach $\la f_2(R) = \la R \approx (\la /\ka) \rho_c \ll 1$, which yields $|\la| \ll 4.24 \times 10^{33}~eV^{-2}$. However, the numerically obtained results show that a slight increase in accuracy would allow an upper bound to be placed on $\la$; this closeness between the prediction of the perturbative model and experiment also seems to validate the latter.

\appendix

\section{}

Given the line element (\ref{line}), one first writes

\beq g^{00}R_{00} - g^{rr}R_{rr} =e^{-\si} { \nu' + \si ' \over r }~~. \eeq

\noindent Resorting to Eq. (\ref{motion}), one gets

\beqa \label{eqapp1} && 2 \ka [2 + a (\rho - 3 p )] (g^{00}R_{00} - g^{rr}R_{rr} ) = \\* \nonumber && 2 (1-ap) (g^{00}T_{00} - g^{rr}T_{rr} ) +4a \ka (g^{00} \square_{00} - g^{rr} \square_{rr})p \rightarrow \\* \nonumber && \ka [2 + a (\rho - 3 p )] e^{-\si} { \nu' + \si ' \over r } = \\* \nonumber && \rho + p -ap(\rho + p ) +4a \ka (e^{-\nu} \square_{00} + e^{-\si} \square_{rr})p ~~, \eeqa

\noindent since the terms depending on the metric cancel out (that is, $g^{00}g_{00}- g^{rr}g_{rr} = 0$).

Next, compute

\beqa \label{eqapp2} && 2 \ka [2 + a (\rho - 3 p )] {R_{\th \th} \over r^2} = \\* \nonumber && (3p - \rho) {g_{\th\th} \over r^2} + 2 (1-ap){T_{\th\th} \over r^2} + {a \ka \over r^2} (4 \square_{\th\th} - g_{\th\th} \square)p \rightarrow \\* \nonumber && 2\ka [2 + a (\rho - 3 p )] \left[ {1\over r^2} + e^{-\si} \left({\si' - \nu' \over 2r} - {1 \over r^2 }\right) \right]= \\* \nonumber && \rho -3 p + 2 (1-ap)p + a \ka\left( {4 \over r^2} \square_{\th\th} + \square\right)p \rightarrow \\* \nonumber && \ka [2 + a (\rho - 3 p )] \left[ {2\over r^2} + e^{-\si} \left({\si' - \nu' \over r} - {2 \over r^2 }\right) \right]= \\* \nonumber && \rho - p - 2 ap^2 + a \ka\left( {4 \over r^2} \square_{\th\th} + \square\right)p ~~. \eeqa

Adding Eqs. (\ref{eqapp1}) to (\ref{eqapp2}), one obtains Eq. (\ref{eq1})

\beqa && \ka [2 + a (\rho - 3 p )] \left[ {1 \over r^2 }+ e^{-\si} \left({\si' \over r} - { 1\over r^2 }\right) \right] = \rho\\* \nonumber && -{ap\over 2}(\rho+3p) + {a \ka \over 2} \left( 5 e^{-\nu} \square_{00} + 3e^{-\si} \square_{rr} + { 2 \over r^2} \square_{\th\th}\right) p~~, \eeqa

\noindent with $\square \equiv e^{-\nu} \square_{00} -e^{-\si}\square_{rr} - 2 \square_{\th\th} / r^2$.

By subtracting Eqs. (\ref{eqapp1}) from (\ref{eqapp2}), and substituting Eq. (\ref{massdef}), one gets Eq. (\ref{eq2})

\beqa && \ka \left[2 + a \left(\rho - 3 p \right) \right] \left[ \left(1 - {2Gm \over r}\right) { \nu' \over r } -{2 G m \over r^3} \right] = p\\* \nonumber && + {a p \over 2} (p-\rho ) + {a \ka \over 2} \left( 3 e^{-\nu} \square_{00} + 5 e^{-\si} \square_{rr} - {2 \over r^2 } \square_{\th\th}\right) p ~~.\eeqa 

\noindent 

\section{}

In order to obtain an approximation to the solution of the Lane-Emden equation (\ref{LE}) in the vicinity of $\xi=0$, one writes $\th(\xi) \approx 1 - A\xi^2 +B \xi^4$. Inserting this in Eq. (\ref{LE}) yields, after some algebra and expanding up to fourth order in $\xi$,

\begin{widetext}\beqa &&\left[ 20 + {A_c \over 4 (n + 1)} \left( A \left[ 4 N_c (n+1) (2n+1) + An (13 + 21n) \right] + 10B (13 + 21 n) \right) \right] \xi^4 = \\* \nonumber && \left[ -1 + A_c {2 + 5 A \over 4} \right] \xi^2+ \left[ An - A_c \left( An \left[ 1 + {11 \over 4} A \right] + {11 \over 2} B \right) \right] \xi^4 ~~.\eeqa \end{widetext}

\noindent Equating second and fourth-order terms, one gets

\beq A \approx { 1 \over 6} \left( 1 - A_c { 13 + 25 n \over 12(n+1) } \right) ~~, \eeq

\noindent and

\beq B \approx {n \over 120} \left ( 1 - A_c {11 \over 60} {39 + 59n \over n+1} \right)~~. \eeq

\section{}

In the unperturbed case, the total mass M is purely gravitational, and hence defined as usual:

\beqa && M = 4 \pi \int_0^{R_r}r^2 \rho ~dr = 4 \pi r_0^3 \rho_{c0} \int_0^{\xi_r}\xi^2 \th_0^n ~d\xi = \\* \nonumber && - 4 \pi r_0^3 \rho_{c0} \int_0^{\xi_r} (\xi^2 \th_0' )' ~d\xi = - 4 \pi r_0^3 \rho_{c0} \xi_r ^2 ( \th_0' )_{\xi = \xi_r} = \\* \nonumber && -4 \pi {R_r^3 \over \xi_r} \rho_{c0} \th_{0r}' ~~, \eeqa

\noindent defining $\th'_{0r} \equiv \th_0'(\xi_r)$; hence, the (unperturbed) central density $\rho_{c0}$, a model dependent parameter, is given by

\beq \rho_{c0} = -{M \over 4 \pi \th_{0r}' } {\xi_r \over R_r^3 }~~.\eeq

In the perturbed case, one first uses Eq. (\ref{massdif}) to define, for clarity,

\beq \Th(\xi) = - {3 n \over 8} {\th_0'^2 \over \th_0} + {\th_0' \over 2 \xi} + {7 \over 8} \th_0^n~~. \eeq

\noindent In the above, $\th_0$ is used instead of $\th$, since $\Th$ is coupled to $A_c$, so that including the perturbation $\de$ would only amount to second-order terms $O(A_c^2)$. Using the definitions $N_c \equiv p_c / \rho_c$ and $r_0^2 \equiv (n+1) p_c / 4 \pi G\rho_c^2 $, one gets

\beqa && M = \int_0^{R_r} m'_e ~dr = \\* \nonumber && 4 \pi \int_0^{R_r} r^2 \rho~dr +{n+1 \over G} A_c N_c \int_0^{R_r} \xi^2 \th_0^n \Th ~dr = \\* \nonumber && 4 \pi r_0^3 \rho_c \int_0^{\xi_r} \xi^2 \th^n ~d\xi +{n+1 \over G} A_c N_c r_0 \int_0^{\xi_r} \xi^2 \th_0^n \Th~d\xi = \\* \nonumber && 4 \pi {R_r^3 \over \xi_r^3} \rho_c \int_0^{\xi_r} \xi^2 \left( \th_0 (1 + A_c \de ) \right)^n ~d\xi + \\* \nonumber && \left( 4 \pi r_0^2 {\rho_c^2 \over p_c} \right) A_c N_c r_0 \int_0^{\xi_r} \xi^2 \th_0^n \Th~d\xi \approx \\* \nonumber&& 4 \pi {R_r^3 \over \xi_r^3} \rho_c \int_0^{\xi_r} \xi^2 \th_0^n \left( 1 + n A_c \de \right) ~d\xi + \\* \nonumber && 4 \pi r_0^3 \rho_c A_c \int_0^{\xi_r} \xi^2 \th_0^n\Th~d\xi = \\* \nonumber&& 4 \pi {R_r^3 \over \xi_r^3} \rho_c \left( \int_0^{\xi_r} \xi^2 \th_0^n ~d\xi + A_c \int_0^{\xi_r} \xi^2 \th_0^n \left(n \de + \Th \right) ~d\xi \right) = \\* \nonumber && 4 \pi {R_r^3 \over \xi_r^3} \rho_c \left( - \xi_r^2 \th_{0r}' + A_c \int_0^{\xi_r} \xi^2 \th_0^n \left(n \de + \Th \right) ~d\xi \right) = \\* \nonumber && -4 \pi {R_r^3 \over \xi_r} \rho_c \th_{0r}' \left( 1 - {A_c \over \xi_r^2 \th_{0r}' } \int_0^{\xi_r} \xi^2 \th_0^n \left(n \de + \Th \right) ~d\xi \right) ~~. \eeqa

\noindent Notice that there are two $A_c$-dependent contributions: one arising from $\th_0$ and its derivatives (embodied in $\Th$), and other involving the integral of $\de$. 

\section{}

By considering the scaling laws $\rho = \rho_c \th^n (\xi)$ and $p = p_c \th^{n+1} ( \xi) $ and using Eq. (\ref{ineqf1a}) one gets, for the $f_1 = f_{1a}$ case,

\beqa && |\la| \gg {2^m\al \ka^{1+m} \over \rho^m p} = \left({2 \ka \over \rho_c }\right)^{m+1} {\al \over 2 N_c \th^{n(1+m)+1}} ~~, \\* \nonumber && |\la| \gg 2^m (1-m) m \al \ka^{1+m} \left|{ \rho'(r) \over \rho^{1+m}p'} \right|= \\* \nonumber && { n \over n+1} \left({2 \ka \over \rho_c} \right)^{m+1} {\al (1-m) m \over 2 N_c \th^{n(1+m) + 1} }~~, \\* \nonumber && |\la | \gg { 2^m (1-m)m \al \ka^{1+m} \over \rho^{1+m} p'' } \left| \rho''(r) -(1+m){\rho'^2(r) \over \rho} \right| = \\* \nonumber && \left({2 \ka \over \rho_c} \right)^{m+1} { \al (1-m)m \over 2 N_c \th^{n(1+m)+1} } \left| {(1+ nm) \th'^2(\xi) - \th \th''(\xi) \over n \th'^2(\xi)+ \th \th''(\xi)} \right| ~~,\eeqa

\noindent while, for the $f_1=f_{1b}$ case, Eq. (\ref{ineqf1b}) yields

\beqa && \la \gg {\al \over 2 N_c \th} ~~, \\* \nonumber && \la \gg {n \over n+1} {\al \over 2 N_c \th} ~~, \\* \nonumber && \la \gg {n \over n+1} {\al \over 2 N_c \th} \left| {(1 - n) \th'^2(\xi) - \th \th''(\xi) \over n \th'^2(\xi)+ \th \th''(\xi)} \right| ~~,\eeqa

\noindent where the prime denotes denotes differentiation with respect to $\xi$ (the factors $r_0$ arising from changing derivation with respect to $r$ to derivation with respect to $\xi = r / r_0$ cancel out).

\begin{acknowledgments}

The authors would like to thank F. Lobo for fruitful discussions.
The work of J.P. is sponsored by the FCT under the grant $BPD 23287/2005$. O.B. acknowledges the partial support of the FCT project $POCI/FIS/56093/2004$.

\end{acknowledgments}

\end{document}